\documentclass[12pt]{article}
\usepackage{amssymb,epsfig,wrapfig,graphics}
\usepackage[dvips]{color}
\usepackage[footnotesize]{caption}

\def\comment#1{{}}

\newlength{\refindent}
\begin{document}
\pagestyle{empty}

\begin{center}
{\Large \bf What Are Gamma-Ray Bursts?\\
\smallskip
The Unique Role of Very High Energy Gamma-Ray Observations\\}
\bigskip\bigskip
{\large
David A. Williams\\
Santa Cruz Institute for Particle Physics\\
University of California, Santa Cruz\\
\smallskip
daw@scipp.ucsc.edu --- (831) 459-3032\\

\bigskip\bigskip
{On behalf of members of the\\ Gamma-Ray Burst Working Group of the study\\
{\it The Status and future of ground-based TeV gamma-ray astronomy}\\
prepared for the Division of Astrophysics of the American Physical Society\\}
}
\bigskip
{\large
D.~A.~Williams, 
A.~D.~Falcone, 
M.~G.~Baring,
J.~Buckley,
V.~Connaughton, 
P.~Coppi, 
C.~Dermer, 
B.~Dingus, 
C.~Fryer, 
N.~Gehrels,
J.~Granot, 
D.~Horan, 
J.~I.~Katz, 
P.~M\'esz\'aros, 
J.~Norris,
P.~Saz~Parkinson, 
A.~Pe'er,  
S.~Razzaque, 
X.-Y.~Wang and 
B.~Zhang
\\
\bigskip
With additional contributions from\\
S. Digel, G. Sinnis and T. C. Weekes\\
\bigskip
}
\end{center}
\newpage
\pagestyle{plain}

\begin{center}
{\bf \large Abstract\\}
\noindent
\begin{minipage}[t]{6.0in}
\noindent
Gamma-ray bursts (GRBs) have been an enigma since their discovery forty years ago.
However, considerable progress unraveling their mysteries has been made in recent years.  Developments
in observations, theory, and instrumentation have prepared the way so that the next decade can be the 
one in which we finally answer the question, ``What are gamma-ray bursts?''  This question encompasses
not only what the progenitors are that produce the GRBs, but also how the enormous luminosity of the GRBs, 
concentrated in gamma rays, is achieved.  Observations across the electromagnetic spectrum, from both 
the ground and space, will be required to fully tackle this important question.  This white paper,
mostly distilled
from a recent study commissioned by the Division of Astrophysics of the American Physical Society,
focuses on what very high energy ($\sim$100 GeV and above) gamma-ray observations can contribute.  
Very high energy gamma rays
probe the most extreme high energy particle populations in the burst environment, testing models of 
lepton and proton acceleration in GRBs and constraining the bulk Lorentz factor and opacity of 
the outflow.  Sensitivity improvements of more than an order of magnitude in the very high energy 
gamma-ray band can be achieved early in the next decade, in order to contribute to this science. 

\end{minipage}
\end{center}

\noindent{\bf \Large Introduction}
\smallskip

Much has been learned since the discovery of gamma-ray bursts (GRBs) in the late 1960s.
There are at least two classes of GRB, most conveniently referred to as ``long'' and ``short,''
based primarily on the duration and spectral hardness of their prompt sub-MeV emission.  
The characteristics of the host galaxies--their types and star formation rates--suggest different 
progenitors for these two classes.
The exact nature of the progenitors 
nevertheless 
remains unknown, although it is widely believed that long GRBs come from the deaths of massive, rapidly rotating stars and short GRBs result from compact object mergers. The unambiguous solution to this mystery is critical to astrophysics since it has fundamental importance to several topics, including stellar formation history and ultra high energy cosmic-ray acceleration. 
Equally important is an understanding of how GRBs process the energy of the explosion into the observed 
gamma-ray luminosity.
A detection of very high energy (VHE, $\sim$100 GeV and above) $\gamma$-ray emission from GRBs would severely constrain the physical parameters pertaining to the particle acceleration by GRBs and the energy injected into the particle acceleration sites.  These observations would constrain models for particle acceleration in bursts,
and the total energy budget could constrain the properties of the GRB progenitors themselves.

The evidence indicates that GRBs involve collimated, relativistic outflows or jets, and one of the big questions is 
whether the jets are dominated by ultrarelativistic protons, which interact with either the radiation field or the
background plasma, or 
are dominated by e$^+$e$^-$ pairs. 
The combination of the {\it Fermi} Large Area Telescope (LAT) and
current generation VHE telescopes will 
contribute to progress on these questions in the near term, but more sensitive observations will be needed
for a full explanation.

The same shocks which are thought to accelerate electrons
responsible for non-thermal $\gamma$-rays in GRBs should also
accelerate protons. The shocks
are expected to be mildly relativistic and to lead to relativistic
protons. The maximum proton energies achievable in GRB shocks are estimated to be 
$\sim$10$^{20}$ eV, comparable to the highest energies of the mysterious ultra high energy cosmic rays measured with large ground arrays. 
The accelerated protons can interact with fireball photons,
leading to 
pions, followed by high energy gamma rays, 
muons, and neutrinos. Photopion production
is enhanced in conditions of high internal photon target density, 
whereas if the density of (higher-energy) photons is too large,
the fireball is optically thick to gamma rays, 
even in a purely leptonic outflow.
High energy gamma-ray studies of GRBs 
provide a direct probe of the 
proton shock acceleration as well as of the photon density.

Most of the information presented in this white paper has been condensed from a more lengthy report \cite{wp}
by the Gamma-Ray Burst Working Group for the study, ``The Status and future of ground-based TeV gamma-ray astronomy,'' completed in 2008 in response to a commission from the Division of Astrophysics of the 
American Physical Society.  Detailed references, provided in that report, are not repeated here.  

\medskip

\noindent{\bf \Large Theoretical Expectations}
\smallskip

Very high energy gamma rays are a natural consequence of many of the processes thought to occur in GRBs.  The
details of the presence or absence, spectrum and light curve of such emission can serve to illuminate which 
processes in fact take place in a particular burst.  A brief discussion of some of the key theoretical ideas
follows.

Gamma-ray burst $\nu F_\nu$ spectral energy distributions peak at photon energies ranging from a few keV to several MeV and are not thermal. From EGRET (and early {\it Fermi}-LAT data \cite{fermi}), it is clear that the 
spectra extend to at least several GeV.
The most widely accepted scenario is the conversion of energy from an explosion into kinetic energy of a
relativistic flow. At a second stage,
the kinetic energy is converted into radiation via internal collisions 
(the internal shock model) resulting from variability
in the ejection from the progenitor or an
external collision (the external shock model) with the surrounding medium. The collisions
produce shocks, which enhance and are believed to create magnetic fields, as well as to accelerate electrons to high energies.  In the standard theoretical model, this initial burst of emission, the so-called prompt emission, is followed by 
afterglow emission from an external shock that moves through the circumburst environment.

An alternative way of producing the emission involves conversion of the explosion energy
into magnetic energy, which
produces a flow that is Poynting-flux dominated. The emission is produced following dissipation
of the magnetic energy via reconnection of the magnetic field lines. 
An apparent advantage of this model over the internal or external shock model is that the conversion of
energy to radiation in principle can be much more efficient. The microphysics of the
reconnection process in this model, like the microphysics determining the 
fraction of energy in relativistic electrons and in the magnetic field
in the internal and external shock scenarios, is not yet fully understood.

VHE observations probe the extremes 
for each of these models 
and simultaneously probe the environment where the emission originated. 

The leading emission mechanism employed to interpret the GRB prompt emission in the keV-MeV band 
is nonthermal synchrotron radiation, although even this is far from
certain forty years after its discovery.
An order of magnitude estimate of the maximum observed energy of photons produced
by synchrotron emission proceeds as follows: Assuming that the
electrons are Fermi accelerated in the shocks, the maximum Lorentz factor of the accelerated
electrons $\gamma_{\max}$ is found by equating the particle acceleration time
and
the synchrotron cooling time,
yielding $\gamma_{\max} = 10^5 / \sqrt{B /10^6}$, where $B$ is the magnetic
field strength in gauss.
For relativistic motion with bulk Lorentz factor $\Gamma$ 
at redshift $z$, synchrotron emission from electrons with $\gamma_{\max}$ 
peaks in the observer's frame at
energy  $70 \, (\Gamma/315) 
(1+z)^{-1}$ GeV,  
which is independent of the magnetic field.
Thus, synchrotron emission can produce photons 
with energies up to, and possibly exceeding, $\sim$100 GeV.

A natural emission mechanism that can contribute to emission at high energies 
($\gtrsim$MeV) is inverse-Compton (IC) scattering. The seed photons for the scattering can be synchrotron photons emitted
by the same electrons, namely synchrotron self-Compton (SSC) emission.
The seed photons can also be  thermal emission originating from the photosphere,
an accretion disk,
an accompanying supernova remnant, or supernova emissions in two-step collapse scenarios.
Compton scattering of photons can produce emission up to observed energies
$15 \, (\gamma_{\max}/10^5)\,(\Gamma/315) 
(1+z)^{-1}$ TeV, 
well into the VHE regime.

The shapes of the Comptonized emission spectra in GRBs depend on the spectra of the seed photons and the energy and pitch-angle distributions of the electrons. A thermal population of electrons can inverse-Compton scatter seed thermal photons or photons at energies below the synchrotron self-absorption frequency to produce the observed peak at sub-MeV energies. Since the electrons cool by the IC process, a variety of spectra can be obtained.
Comptonization can produce a dominant high energy component 
that can explain hard high-energy spectral components, such as that observed in
GRB 941017. 
Prolonged higher energy emission could potentially be observed with a sensitive VHE gamma ray instrument.

The maximum observed photon energy
from GRBs 
is limited by the annihilation of  gamma rays with target photons, both extragalactic IR background and photons local to the GRB, to produce electron-positron pairs. This limit is sensitive to the uncertain value of the bulk motion Lorentz factor as well as to the spectrum at low energies, and is typically in the sub-TeV regime.
Generally, escape of high-energy photons requires large Lorentz factors.
In fact, observations of GeV photons have been used to constrain the minimum Lorentz factor of the bulk motion of the flow, and spectral 
coverage up to VHE energies could 
further constrain
the Lorentz factor.
If the Lorentz factor can be determined independently, {\it e.g.}\ from afterglow modeling,
then the annihilation signature can be used to diagnose the gamma-ray emission 
region.

The evidence for acceleration of leptons in GRB blast waves is based on fitting lepton synchrotron spectra models to GRB spectra. This consistency of leptonic models with observed spectra still allows the possibility of hadronic components in these bursts, and perhaps more importantly, GRBs with higher energy 
emission
have not been explored for such hadronic components due to the lack of sensitive instruments in the GeV/TeV energy range. The crucially important high-energy emission components, represented by a few EGRET bursts, 
a marginal significance Milagrito TeV detection, 
and now GRB\,080916C \cite{fermi}, 
have been statistically inadequate to look for 
correlations between high-energy and keV/MeV emission
that can be attributed to a particular 
process. Indeed, the 
prolonged
high-energy components in GRB\,940217 and the ``Superbowl'' burst, GRB\,930131, and the anomalous gamma-ray emission component in GRB\,941017, behave quite differently than the measured low-energy gamma-ray light curves. Therefore, it is quite plausible that hadronic emission components are found in the high energy spectra of GRBs.

Several theoretical mechanisms exist for VHE $\gamma$-ray emission from hadrons. 
Accelerated protons can emit synchrotron radiation in the GeV--TeV energy band. The power emitted by a particle is 
$\propto \gamma^2/m^2$, where $\gamma$ is the Lorentz factor of the particle and $m$ is its mass. 
Given the larger mass of the proton, 
to achieve the same output luminosity, 
the protons have 
$\sim$1836 times 
higher mean Lorentz factor, 
the acceleration mechanism must convert 
$\sim$ 3 million times 
more energy to protons than electrons 
and the peak of the proton emission would be at 
$\gtrsim$ 2000 times higher  
energy 
than the peak energy of photons emitted by the electrons.
Alternatively,
high-energy baryons can produce energetic pions, via photomeson interactions with low energy photons, creating 
high-energy photons and neutrinos as the pions decay. This process could be the primary source of ultra high energy (UHE)
neutrinos. 
Correlations between gamma-ray opacity and bulk Lorentz factor from measurements 
with sensitive VHE $\gamma$-ray telescopes can test for UHE cosmic-ray production. 
Finally, 
proton-proton or proton-neutron collisions may also be a
source of pions, and in addition, 
if there are neutrons in the flow, then the neutron $\beta$-decay has a drag effect on the protons, which may produce another source of radiation. Each of 
these cases has a VHE 
spectral shape and intensity 
that can be studied coupled with the 
emission measured at lower energies and with neutrino measurements.

Afterglow emission is explained in synchrotron-shock models by the same processes that occur during the prompt phase.  However, unlike the prompt phase, there are optical and near infrared linear polarization measurements to support the synchrotron 
interpretation.  The key difference is that the afterglow emission originates from large radii,
$\gtrsim$10$^{17}$~cm, as opposed to the much smaller radius of the flow during the prompt emission
phase,  $\sim$10$^{12} - 10^{14}$~cm for internal shocks, and $\sim$10$^{14} - 10^{16}$~cm for external shocks. As a result, the density of the blast-wave shell material is smaller during the afterglow emission phase than in the prompt phase, and some of the radiative mechanisms, e.g. thermal collision processes, may become less important.
The lower density also reduces the expected photon pair opacity, enabling VHE gamma rays to more readily 
escape, if produced.

\medskip

\noindent{\bf \Large Status of Observations}
\smallskip

One definitive observation
of prompt or afterglow VHE $\gamma$-ray emission could significantly improve our
understanding of the processes at work in the GRB and its
aftermath. Although many authors have predicted such emission, the predictions
are near or below the sensitivity of current instruments, and there
has been no definitive detection of VHE emission from a GRB either
during the prompt phase or at any time during the afterglow.  The detection of
a 13 GeV photon by the {\it Fermi}-LAT from GRB\, 080916C \cite{fermi} at 
a redshift of 4.35$\pm$0.15 \cite{grond} shows that photons with at least 70 GeV
can be produced by GRBs. 

For the observation of photons of energies above 300\,GeV, only
ground-based telescopes are suitable. These ground-based
telescopes fall into two broad categories, air shower arrays and
imaging atmospheric Cherenkov telescopes (IACTs). The air shower arrays,
which have wide fields of view that are suitable 
for GRB searches, are relatively less sensitive. There are several reports
from these instruments of possible TeV emission: 
emission $>$16 TeV from GRB\,920925c,
an indication of 10\,TeV emission in a stacked analysis of 57 bursts,
and
an excess gamma-ray signal during the prompt phase of
GRB\,970417a. In all of
these cases however, the statistical significance of the detection is
not high enough to be conclusive. 
In addition to searching the Milagro 
data for VHE counterparts for over 100 satellite-triggered GRBs since 2000,
the Milagro Collaboration
conducted a search for VHE transients of 40 seconds to 3 hours
duration in the northern sky; no evidence
for VHE emission was found from these searches.

IACTs have better flux sensitivity and energy resolution 
than air shower arrays,
but are limited by their
small fields of view (3--5$^\circ$), which require retargeting to the 
GRB position, and low duty cycle ($\sim$10\%).
Follow-up observations on many GRBs have been 
made by the MAGIC, VERITAS, and HESS groups, with upper limits typically 
around 5\%\ of the Crab flux or better.
MAGIC observations of GRB\,050713a began only 40 seconds after the prompt
emission, demonstrating the fast slewing capability of that instrument.

One of the main obstacles to VHE detection of GRBs is their distance.
Pair production interactions of gamma rays with the infrared
photons of the extragalactic background light attenuate the gamma-ray
signal, limiting the distance over which VHE gamma rays can
propagate.  The MAGIC Collaboration has reported the VHE detection of the flat
spectrum radio quasar 3C279, at 
redshift 0.536. This represents a large increase in distance
to the furthest detected VHE source, revealing more of the
universe to be visible to VHE astronomers than was previously thought.
\medskip

\begin{figure*}[!b]
\centering
\includegraphics[width=5in]{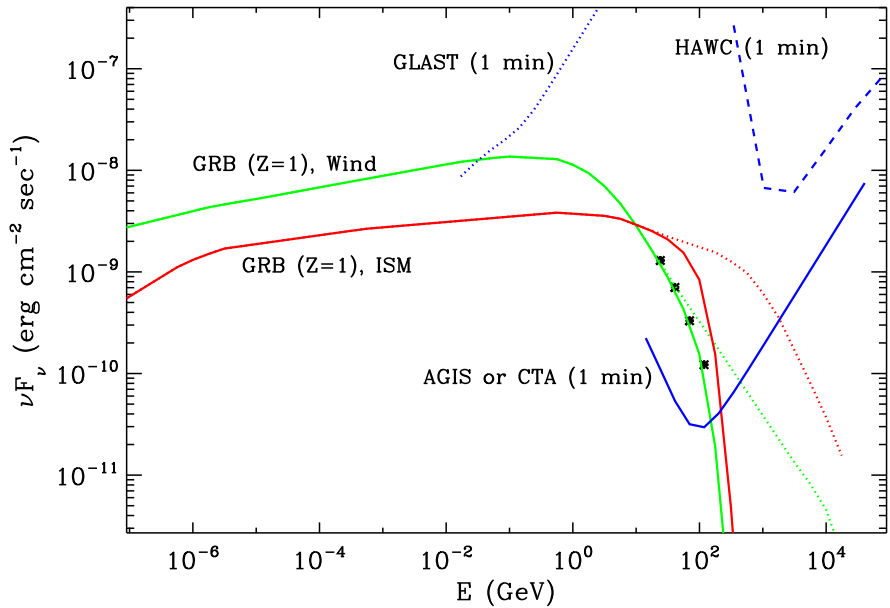}
\caption{
The sensitivity of high-energy gamma-ray instruments for 1 minute observations of an object in the field of view.
AGIS, CTA and HAWC are concepts for state-of-the-art instruments which could be realized in the next decade and 
which improve upon the sensitivity of current instruments of the same type by an order of magnitude or more.  AGIS and CTA are narrow-field ($\sim$5--10$^\circ$) Cherenkov telescope arrays which operate at night; HAWC is a 
wide-field ($\sim$2 sr) air shower array which operates around the clock.  
The AGIS/CTA curve (solid blue) is the differential sensitivity for 
0.25 decade bins, while for HAWC (dashed blue)
0.5 decade bins are assumed.  The dotted blue curve shows the sensitivity of the {\it Fermi}-LAT (``GLAST'').
The sensitivity curves are based on a 5$\sigma$ detection and at least 25 detected photons in each bin.
Also shown are plots of the predicted gamma-ray spectra for a GRB at a redshift of z=1 adapted from Pe'er and Waxman 
\cite{PeerWaxman05b},
reduced by a factor of 10 to illustrate the sensitivity even to 
weaker bursts.
The green and red curves show the calculation for a wind environment and an ISM-like environment. The dotted curves give the source spectrum, while the solid curves include the effects of extragalactic background light absorption using the model of Franceschini {\it et al.}\ \cite{Franceschinietal2008}.
Black points and error bars (not visible) are simulated independent spectral points that could be obtained with AGIS/CTA.}
\label{grb_sedsim}
\end{figure*}

\noindent{\bf \Large Observations in the Next Decade}
\smallskip

Ground-based observations of VHE emission from GRBs are difficult.
Nevertheless, instruments planned for the next decade 
likely will be able to detect the VHE component of GRBs,
which could open a new way to understand as well as discover them.  
The fraction of bursts close enough to 
elude attenuation at VHE energies by the extragalactic background light 
is small. Only $\sim$10\% of long bursts have redshift $<$0.5, 
the redshift of the most distant detected VHE 
source.  Roughly half of short burst redshifts are $<$0.5, but the prompt emission has ended 
prior to satellite notifications of the burst location.  
The reach of VHE instruments can be
extended by lowering the energy threshold, since the attenuation is a rapidly rising function 
of energy.
The GRBs that will be detected from the ground will preferentially be nearby,
making them
excellent candidates for multiwavelength and multimessenger ({\it e.g.}\ LIGO 
for neutron star mergers or IceCube for neutrino production) observations.

Wide field of view detectors with high duty cycle operations 
would be ideal to observe the prompt emission from 
gamma-ray bursts.  Imaging atmospheric Cherenkov telescopes (IACT) can be made 
to cover large sections of the sky by either having many 
mirrors each pointing in a separate direction or by employing secondary optics to expand the field of view of each mirror. 
However, the duty factor is still $\sim$10\% 
due to solar, lunar, and weather constraints. IACTs could also be made with fast slewing mounts to allow them to slew to most GRBs within $\sim$20 seconds, thus allowing them to observe some GRBs before the end of the prompt phase. 
Alternatively, extensive air shower 
detectors intrinsically have a field of view of $\sim$2 sr and operate with $\sim$95\% 
duty factor. These observatories, especially if located at very high altitudes, 
will have $\sim$100 m$^2$ of sensitive area at 100 GeV and be more sensitive than {\it Fermi}-LAT above
100 GeV, but with poor energy resolution.
IACT arrays will be able to achieve a sensitivity of 10$^{-9}$ erg s$^{-1}$ cm$^{-2}$ (in $\lesssim20$ sec integration), or better, sufficient to probe a substantial fraction of models for high-energy emission from those
bursts which they are able to observe.
The combined observations of both of these types of detectors would yield the most complete picture of the prompt high energy emission. 

Observational VHE $\gamma$-ray astrophysics is young, and substantial headway has been made in
recent years understanding the performance and limitations of present instruments.  As a result,
concepts for next-generation instruments of both the extensive air shower and imaging atmospheric
Cherenkov telescope design are at hand and could be realized early in the next decade.  In both 
cases, improvements of more than an order of magnitude in sensitivity are projected. 
The expected sensitivity of the two techniques 
is shown in Figure~\ref{grb_sedsim}.

The detector strategy for extended emission associated with traditional afterglows or with late-time flares from GRBs is far simpler than the strategy for early prompt emission. The high sensitivity and low energy threshold of an IACT array are the best way to capture photons from this emission at times greater than $\sim$1 min, particularly if fast slewing is included in the design.

\medskip

\noindent{\bf \Large Summary}
\smallskip

Gamma-ray bursts undoubtedly involve a population of high-energy 
particles responsible for the emission detected
from all bursts (by definition) at energies up to of order 1 MeV,
and for a few bursts so far observable by EGRET and {\it Fermi}-LAT, up to a few GeV.  Gamma-ray 
bursts may in fact be the source of the highest energy particles in the universe.
In virtually all models, this high-energy population can also produce
VHE gamma-rays, although in many cases the burst environment would be 
optically thick to their escape.  The search for and study of VHE emission from GRBs therefore tests theories about the nature of these
high energy particles (Are they electrons or protons? What is their
spectrum?) and their environment (What are the density and bulk
Lorentz factors of the material? What are the radiation fields?
What is the distance of the emission site from the central source?).
In addition, sensitive VHE measurements would aid in assessing the 
the total bolometric radiation output from bursts.  
Knowledge of the VHE gamma-ray properties of bursts will therefore
help complete the picture of these most powerful known accelerators.

An example of the insight that can be gleaned from VHE data
is that leptonic synchrotron/SSC
models can be tested, and model parameters extracted, by 
correlating the peak energy of
X-ray/soft $\gamma$-ray emission
with GeV--TeV data.
For long-lived GRBs, the spectral properties of late-time flaring in the X-ray band can be compared to the measurements in the VHE band, where associated emission is expected.
Of clear interest is whether there are distinctly evolving high-energy $\gamma$-ray  
spectral components, whether at MeV, GeV or TeV energies, unaccompanied
by the 
associated lower-energy component 
expected in leptonic
synchro-Compton models. 
Emission of this sort is most easily explained in models involving
proton acceleration.
As a final example, the escape of VHE photons from the burst fireball provides a tracer of the minimum Doppler boost and bulk Lorentz motion of the emission region 
along the line of site, since the inferred opacity of the emission region declines with increasing boost.   

There are observational challenges for detecting VHE emission during the initial
prompt phase of the burst.  The short duration of
emission leaves little time (tens of seconds) for repointing 
an instrument, and the opacity of the compact fireball is at its 
highest.  For the majority of bursts having redshift $\gtrsim$0.5, the 
absorption of gamma rays during all phases of the burst by 
collisions with the extragalactic background light reduces the 
detectable emission, more severely with increasing gamma-ray energy.
With sufficient sensitivity, an all-sky
instrument is the most desirable for studying the prompt phase,
in order to measure the largest sample of bursts and to catch 
them at the earliest times.
The techniques used to implement all-sky compared to 
pointed VHE instruments result in a trade-off of energy threshold and
instantaneous sensitivity for field of view.  More than an order of magnitude improvement in sensitivity to GRBs is possible now
for the next-generation instruments of both types, giving both 
approaches a role in future studies of GRB prompt emission.

The detection of VHE afterglow emission, delayed prompt emission from large radii, and/or late X-ray flare-associated emission simply requires a sensitive instrument 
with only moderate slew speed. It is likely that an instrument with significant sensitivity improvements over the current generation of IACTs will detect GRB-related VHE emission from one or all of these mechanisms which do not suffer from high internal absorption, thus making great strides towards understanding the extreme nature and environments of GRBs and their ability to accelerate particles.

In conclusion, large steps in understanding
gamma-ray bursts have frequently resulted from particular new 
characteristics measured for the first time in a single burst. 
New instruments improving sensitivity to VHE $\gamma$-rays
by an order or magnitude or more compared to existing observations
have the promise to make just
such a breakthrough in the VHE band during the next decade, helping to 
answer the question, ``What are gamma-ray bursts?''


\vspace{-2.8ex}
\begin{thebibliography}{}


\setlength{\itemsep}{-0.0em}
\setlength{\parsep}{4pt}
\setlength{\baselineskip}{12pt}

\bibitem{wp} A. D. Falcone {\it et al.} 2008, arXiv:0810.0520, http://cherenkov.physics.iastate.edu/wp/grb.html

\bibitem{fermi} F. Piron 2008, Sixth Huntsville Gamma-Ray Burst Symposium; A. Abdo {\it et al.} 2009, Science, in press.

\bibitem{grond} J. Greiner {\it et al.} 2009, submitted to Astron.\ \& Astrophys., arXiv:0902.0761

\bibitem{PeerWaxman05b}  A. Pe'er \& E. Waxman 2005, Astrophys.\ J., 633, 1018

\bibitem{Franceschinietal2008} A. Franceschini, G. Rodighiero \& M. Vaccari 2008, Astron.\ \& Astrophys., 487, 837

\end{thebibliography}
\end{document}